\documentclass[aps,prl,floats,twocolumn,showpacs,superscriptaddress]{revtex4}

\usepackage{graphicx,epsfig}
\usepackage{times}
\usepackage{graphics,dcolumn,bm,fleqn,epic,eepic,float}
\usepackage{amssymb,amsmath,multirow,rotate,color}
\begin{document}

\title{Scaling breakdown in flow fluctuations on complex networks}
\author{Sandro Meloni}
\affiliation{Department of Informatics and Automation, University of Rome "Roma Tre", Via della Vasca Navale, 79
00146, Rome, Italy}

\author{Jes\'us G\'omez-Garde\~nes}
\affiliation{Scuola Superiore di Catania, 
Via San Nullo 5/i, 95123 Catania, Italy}\affiliation{Institute for Biocomputation and Physics of Complex
Systems (BIFI), University of Zaragoza, 50009 Zaragoza, Spain}

\author{Vito Latora}
\affiliation{Dipartimento di Fisica e
    Astronomia, Universit\`a di Catania, and INFN, Via S. Sofia 64, 95123
    Catania, Italy}
  
\author{Yamir Moreno}\thanks{E-mail: yamir@unizar.es}

\affiliation{Institute for Biocomputation and Physics of Complex
Systems (BIFI), University of Zaragoza, 50009 Zaragoza, Spain}

\affiliation{Department of Theoretical Physics, University of Zaragoza, 50009 Zaragoza, Spain}

\date{\today}

\begin{abstract}
We propose a model of random diffusion to investigate flow
fluctuations in complex networks. We derive an analytical law
showing that the dependence of fluctuations with the mean traffic
in a network is ruled by the delicate interplay of three factors,
respectively of dynamical, topological and statistical nature. 
In particular, we demonstrate that the
existence of a power-law scaling  characterizing the flow fluctuations at 
every node in the network can not be claimed
for. We show the validity of this scaling breakdown under quite general topological and dynamical situations by means of different traffic algorithms and by analyzing real data.
\end{abstract}

\maketitle

Communication networks \cite{book1} are nowadays the subject of
intense research as modern society increasingly depends on them. On the  
one hand, the first studies have dealt with the architecture of these 
systems, showing that the systems' topological features
\cite{book1,pvv01,wms06} are at the root of the critical behavior of
several dynamical processes taking place on top of them
\cite{book1,physrep}. On the other hand, models for traffic and
information flow on complex networks have been recently investigated as
a way to improve our understanding on key issues such as the
scalability, robustness, performance and dynamics of technological 
networks \cite{book1,physrep}.  In particular, much effort has been
invested in finding what are the conditions for an efficient
performance of communication networks, the latter being measured as
the ability of the system to avoid congestion and reduce transit times
\cite{gdvca02,egm04,egm05,sclts07}. Nevertheless, large
communication networks such as the Internet usually avoid the regime
in which congestion arises and therefore the dynamics of packets is
not driven by congestion processes. Instead, the fluctuations in
traffic flow constitute the main factor affecting the dynamics of
these communication systems.

The relationship between the fluctuations $\sigma$ and the average
flux $\langle f \rangle$ in traffic dynamics on complex networks is a 
controversial issue that has received a lot of attention very recently 
\cite{flowba,flowalex}.  
The authors of Refs.\ \cite{flowba,flowba2} claimed the existence of the 
relation $\sigma \sim \langle f \rangle^{\alpha}$, with   
real communication networks belonging to one of two universality 
classes, the first one characterized by an exponent value $\alpha=1/2$, 
the second one by $\alpha=1$. 
The authors of \cite{flowalex} questioned the existence of
the two universality classes. They numerically showed that there is a wide spectrum of possible values for $\alpha$, depending on parameters such as the persistence of packets in the network, the duration of the time window during which statistics are recorded, and the rate of service at the nodes' queues \cite{flowalex}.

In this Letter, we propose a model for traffic in complex networks, the
Random Diffusion (RD) model, that is amenable to analytical solution.
The model predicts the existence of a simple law that relates the
fluctuations at a node $i$, $\sigma_i$ to the average traffic flow
$f_i$, depending on the delicate balance of three quantities: (i) the
variation in the number of packets in the network, (ii) the degree of
the node $i$, and (iii) the length of the time window in which
measures of traffic flow are performed.  Notwithstanding its
simplicity, the RD model is able to capture the essential ingredients
determining the scaling of fluctuations empirically observed for
traffic flow in real complex networks.  More important, we also show
that the hypothesis of a {\it power-law scaling} of flow fluctuations 
has to be abandoned under certain conditions. Results of numerical
simulations of a traffic-aware model and analysis of real data of
Internet flow confirm our theoretical findings.

In the random diffusion (RD) model we represent packets of 
information as $w$ random walkers traveling in a network made up of $N$ nodes and $K$ links.   
Under the assumption that the packets are not interacting, it follows 
that the average number of walkers $\lambda_i$ at a node $i$ 
is given, in the stationary regime, by \cite{nr04,gl07}
\begin{equation}
  \lambda_i (w)= \frac{k_i}{2K}w\;.
\end{equation}
Let us assume that the total observation time $T$ is divided into
time-windows of equal length. Each window is made of $M$ time units.
A window represents the minimal resolution for measurements of the
flux in a node and its fluctuations, being the first the result of
accumulating the number of packets traveling through the node during
the $M$ time units. The average number of packets $\langle f_i
\rangle$ processed by node $i$ in a time window is measured, together
with its standard deviation $\sigma_i$.  These are the two quantities
monitored in Refs.\cite{flowba,flowalex} for real systems and in
the numerical simulations of network traffic models.  The main interest
is to investigate the dependence of $\sigma_i$ with $\langle f_i \rangle$. 
In particular, we want to verify whether a power-law relation 
$\sigma_i \sim \langle f_i \rangle^{\alpha}$ holds, and what factors 
determine the exponent $\alpha$. In the RD model we can consider 
two possible situations: either the number of packets in the network 
is constant over the whole period of time $T$,  
namely $w=W$, or it can vary from one time window to
the other. In the latter situation, we assume that
the probability $F(w)$ of having $w$ walkers on the network in a 
window of length $M$ is equally distributed in the range 
$[W-\delta, W+\delta]$, i.e., 
\begin{equation}
  F(w)=\frac{1}{2\delta+1}\;,
\label{pw} 
\end{equation}
with $ 1 \le \delta \le W$. 
To find an expression for the average number of packets $\langle f_i \rangle$  
flowing through a given node $i$, we first calculate the probability  
$P_i(n)$ that, after $M$ time steps, $n$ packets have visited node $i$.  

In the case $w=W$,  due to the fact that the packets are not interacting,  
the arrival of walkers at a node is a Poisson process. 
Therefore, after a period of $M$ time units, the
mean number of packets (the average flux) at a node $i$ is $\langle
f_i\rangle=\lambda_i (w)M$, and the probability of having $n$ packets 
reads
\begin{equation}
P_i(n)={\mbox e}^{-\lambda_i (w) M}\frac{(\lambda_i (w)M)^n}{n!}\;, 
\label{eq:Poisson}
\end{equation}
with $\sigma=\sqrt{\lambda_i (w) M}=\sqrt{\langle f_i\rangle}$. 
Thus, the scaling exponent is $\alpha=1/2$.

In the more general case in which the number $w$ is distributed as in Eq.~(\ref{pw}),  
the probablity $P_i(n)$ is
\begin{equation}
  P_i(n)=\sum_{j=0}^{j=2\delta}\frac{{\mbox e}^{-\frac{k_i}{2K}(W-\delta+j)M}}{2\delta+1} 
\frac{ [\frac{k_i}{2K}(W-\delta+j)M ]^n}{n!}\;.
\label{eq:var}
\end{equation}
Calculating first and second moments of $P_i(n)$ one obtains 
\begin{eqnarray}
\langle f_i\rangle &=& \sum_{n=0}^{\infty}nP_i(n) = \frac{k_i W M}{2K}\;,
\\
\langle f_i ^2\rangle &=& \sum_{n=0}^{\infty}n^2P_i(n) = 
\langle f_i\rangle^2 (1 + \frac{\delta^2}{W^2}) +\langle f_i\rangle\;.
\end{eqnarray}
Finally, the standard deviation can be expressed 
as a function of $\langle f_i\rangle$ as 
\begin{equation}
 \sigma_{i}^2 = \langle f_i\rangle\left(1+\langle f_i\rangle\frac{\delta^2}{W^2}\right)\;.
\label{sd}
\end{equation}

The above derivation provides an understanding of the origins of Eq.\ (\ref{sd}), proposed in \cite{flowba},
and shows that the relation between $\sigma_{i}$ and $\langle f_i \rangle$ 
depends on the concurrent effects of three factors, namely:  
{\it (i)} the noise $\delta$ associated to the fluctuations in 
the number of packets
in the network from time window to time window;
{\it (ii)} the length $M$ of the time window;  and
{\it (iii)} the degree of the node $k_i$ (since $\langle f_i\rangle$ depends 
on $k_i$). Consequently, real traffic rarely falls in either of the two limiting cases of Eq.\ (\ref{sd}), i.e., $\sigma\sim \langle f\rangle^{\alpha}$ with $\alpha=1/2$ or $1$.

Expression (\ref{sd}) contains all the behaviors 
previously observed in Refs.~\cite{flowba,flowalex}, and also predicts 
new dependencies that can be tested to be valid in more refined 
traffic models as well as in real data. 
In fact, if the three quantities $\delta$, $M$ and $k_i$ are such that
\begin{equation}
 \frac{k_i M \delta^2}{2KW}\ll 1 \;,
\label{ratio}
\end{equation}
expression (\ref{sd}) reduces to a power-law scaling 
$\sigma\sim \langle f\rangle^{\alpha}$ with exponent 
$\alpha=1/2$. On the contrary, whenever the ratio 
$\frac{k_i M \delta^2}{2KW}$ is not negligible anymore, the exponent 
$\alpha$ differs from $1/2$ and approaches 1. In other words, it may well be
the case in which, even for small values of the noise parameter
$\delta$, a large value of $M$ cancels out the effect of the ratio
$\frac{\delta}{W}$ being too small in Eq.\ (\ref{sd}). This behavior was already explored in \cite{flowalex} by means of numerical simulations. However, the fact that the ratio in formula (\ref{ratio}) depends quadratically on 
$\delta$ and only linearly on $k_i$ and $M$, has gone unnoticed. The RD model puts such dependence on 
solid theoretical grounds, and also reveals the role played by the 
other two parameters $M$ and $k_i$ on the observed scaling. 
\begin{figure}[!t]
\begin{center}
\epsfig{file=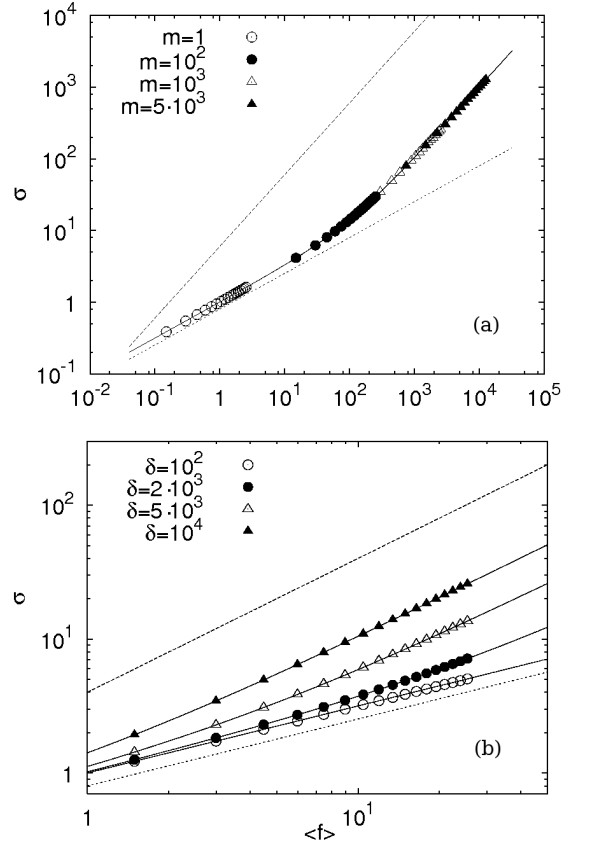,width=\columnwidth,angle=0,clip=1}
\end{center}
\caption{Flow fluctuation $\sigma$ as a function of $\langle f\rangle$ for the RD model
  with various parameter values. In panel (a), $\delta=10^3$ and
  $W=10^4$. In panel (b), $W$ has the same value while $M$ has been
  fixed to 10. In both figures, points correspond to the solution of
  Eq.\ (\ref{sd}) for different values of $k_i$ ($1\ldots 18$). The
  total number of links is $K=33500$. Dashed lines are guides to the
  eyes and correspond to $\sigma\sim\langle f\rangle^{\alpha}$, with
  $\alpha=1/2$ (lower curves) and $\alpha=1$ (upper curves).  See the
  text for further details.}
\label{fig1}
\end{figure}

In Fig.\ \ref{fig1} we plot the dependence of $\sigma$ with $\langle f 
\rangle$ in the RD model for several values of the parameters $M$ and $\delta$. Panel
(a) corresponds to the case in which the ratio
$\frac{\delta}{W}=10^{-1}$ is fixed and the length of the time windows
used to measure the flow of packets through different nodes is
varied. For each value of $M$, we have superimposed the results obtained for
nodes with different connectivity values, ranging from $k_i=1$ to
$k_i=18$.  If one follows the arguments given in \cite{flowba}, a
value of $\alpha=1/2$ should be expected for this choice of
$\delta/W$. Instead, as shown in the figure, $\sigma\sim
\langle f\rangle^{\frac{1}{2}}$ only for small values of $M$, while
the scaling exponent approaches $1$ as $M$ is increased. This means
that, whenever the temporal resolution in the measurements is not small
enough and packets are counted and accumulated over long periods, $\alpha$ tends to $1$. 

A novel striking feature revealed by law (\ref{sd}), and 
not revealed in previous studies, is the dependence with
the degree of the nodes.
An example of the effects of node degrees is shown in Fig.\ \ref{fig1}{\em a}. 
It turns out that, for some values of $M$ (e.g. $M=10^2$ in the figure),  
the fluctuations at lowly connected nodes are
characterized by an exponent $\alpha=1/2$, whereas for highly
connected nodes the exponent turns out to be $\alpha=1$. Hence, there
is not a single exponent characterizing the fluctuations at {\em every}
node of the network, regardless of its connectivity.
This is again a clear indication that a power-law behaviour,  
$\sigma\sim \langle f\rangle^{\alpha}$, even with non-universal 
exponents ranging in $[1/2,1]$, is not the most general situation 
when characterizing the flow fluctuations for a whole network \cite{flowba,flowalex}.
Admittedly, $\alpha$ is not constant for every possible choice of the parameters
$\delta$, $W$ and $M$ along the whole set of $k_i$ values. This effect is
particularly relevant for highly heterogeneous networks like the
Internet, where degree classes span several decades. In these kinds of
networks, one should therefore expect different scaling laws depending
on whether the packets are flowing through lowly or highly connected
nodes.

The influence of the noise level on $\alpha$ for a fixed time window
length ($M=10$) is depicted in Fig.\ \ref{fig1}{\em b}. When $\delta$
is small, so that the number of packets in the network from one time
frame to the following does not change significantly, $\alpha=1/2$. On
the contrary, when $\delta$ is sufficiently large, the exponent is
$1$. This is more in consonance with the results in \cite{flowba},
where the dependence with the noise level was addressed only for a low value of $M$, getting that as $\delta$
increases $\alpha\rightarrow1$. On the other hand, we observe again
that fixing $M$ and varying $\delta$ does not guarantee the existence
of a unique exponent for the scaling of fluctuations in traffic flow,
though in this case the dependence is smoother than that observed in
Fig.\ \ref{fig1}{\em a}.

\begin{figure}[!t]
\begin{center}
\epsfig{file=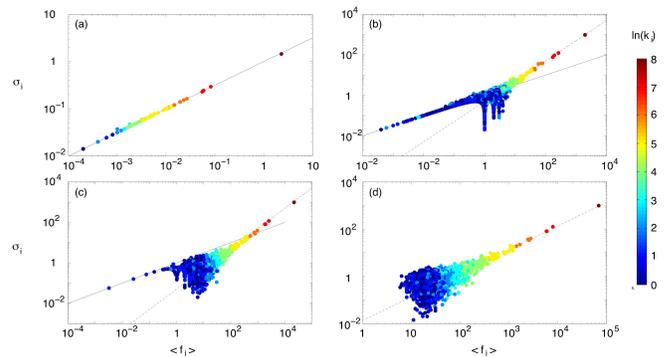,width=\columnwidth,angle=0,clip=1}
\end{center}
\caption{(color online) Flow fluctuation $\sigma$ as a function of $\langle f\rangle$ from 
numerical simulations of the Internet traffic model (see text for details) 
on synthetic scale-free networks with $N=10^4$ nodes, $K=37551$ links, 
and degree exponent $\gamma=2.2$. 
Different panels correspond to different values of $M$, 
respectively $M=1,5\times 10^2, 35\times 10^3,10^5$. Color-coded values represent the logarithm 
of node degree. The continuous line is the curve $y=x^{0.5}$, while 
the dashed line is $y\sim x$.} 
\label{fig2}
\end{figure}

In the following we show that expression (\ref{sd}) predicted by
the RD model is indeed valid for more elaborated traffic models, 
and that the RD approximation
captures the phenomenology of real communication systems.  We report the results obtained on top of synthetic scale-free (SF) networks with $N=10^4$ nodes and power-law degree distributions $p_k \sim k^{-\gamma}$, with an
exponent $\gamma=2.2$ as the one empirically observed for the Internet
at the autonomous system level \cite{pvv01}. However, we stress that since the topological properties of the underlying graph only enter into Eq.\ (\ref{sd}) through the degree of the nodes $k_i$ and the total number of links in the network, $K$, the results hold for any graph with an arbitrary degree distribution $p_k$ as our own simulations using SF networks, random graphs and a real autonomous system map of the Internet \cite{pvv01} reveal.

On the other hand, to mimic the way packets flow in real communication networks, we consider a dynamical model that is able to simulate Internet's
most important dynamical characteristics \cite{egm04,egm05}. The dynamics of the
packets is simulated as follows. Each node represents a router with an
infinite size buffer. The delivery of packets is made following a
First In First Out (FIFO) policy. At each time step, $p$ new packets are introduced in the system with randomly chosen sources and
destinations \cite{note1}. Packets routing
is based on a traffic-aware scheme \cite{egm04,egm05} in which the
path followed by a packet is that that minimizes the effective
distance $d_{\text{eff}}^i = h d_i+(1-h)c_i$, where $d_i$ is the
distance between node $i$ and the packet destination, $c_i$ is the
number of packets in $i$'s queue, and $h$ is a tunable parameter that
accounts for the degree of traffic awareness incorporated in the
delivery algorithm \cite{egm04,egm05}. 
It is worth recalling that $h=1$ recovers a shortest-path delivery protocol, mimicking most of the actual Internet routing mechanisms. 

\begin{figure}[!t]
\begin{center}
\epsfig{file=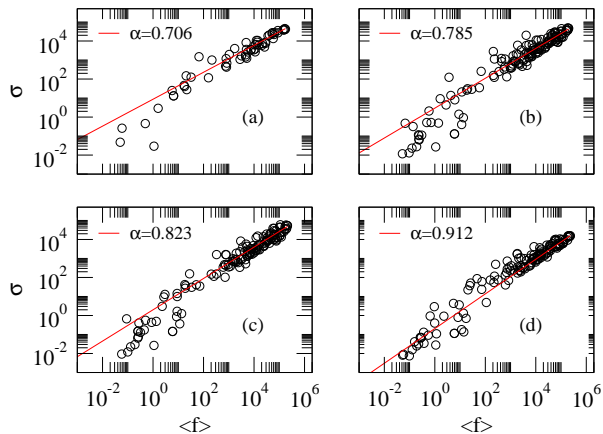,width=0.8\columnwidth,angle=-90,clip=1}
\end{center}
\caption{Flow fluctuation $\sigma$ as a function of $\langle f\rangle$ for
  the Abilene Interfaces. The values of $M$ used in each panel are:
  $M=5$ (a), $M=30$ (b), $M=60$ (c), and $M=720$ (d). Time is in
  minutes. The value of $\alpha$ for each $M$ is also 
  reported. Averages are taken over one month of data corresponding to
  the period between January 11 to February 11 of 2006.}
\label{fig3}
\end{figure}

Figure\ \ref{fig2} shows $\sigma$ as a function of $\langle f\rangle$
obtained through extensive numerical simulations of the traffic model
with $h=1$ and $p=2$.  Different panels in the figure correspond to
different values of the time-window length $M$. The results indicate
that the main responsible of the value of $\alpha$ (interpolating
between the two extreme $\alpha=1/2$ and $\alpha=1$) is the interplay
between the node degree and the time resolution used to record the
flux of packets, exactly as predicted by the scaling law (\ref{sd})
obtained in the RD model.  In fact, Fig.\ \ref{fig2}{\it a}
corresponds to the choice of parameters for which formula
(\ref{ratio}) holds for all values of $k_i$, leading to
$\alpha=1/2$. On the contrary, when $M$ is large enough and the other
parameters are kept fixed as in Fig.\ \ref{fig2}{\it d}, relation
(\ref{ratio}) is not satisfied whatever the value of $k_i$ used, hence
giving an exponent $\alpha=1$. Finally, the breakdown of the scaling
law $\sigma\sim\langle f\rangle^{\alpha}$ anticipated by the RD model
is captured in Figs.\ \ref{fig2}{\it b} and {\it c}, where it is
clearly revealed that there is not a unique exponent characterizing
the flow of packets through {\em every} node of the network. Indeed,
there is a crossover from $\sigma\sim\langle f\rangle^{1/2}$ for lowly
connected nodes to $\sigma\sim\langle f\rangle$ for the highly
connected ones. We also note that a similar behavior is observed 
(figures not shown) when traffic-aware routings ($h<1$) are taken into 
account.

Finally, we have also analyzed the data corresponding to the traffic
between routers of the Abilene backbone network \cite{abilene}. As the data collected for the routers in the backbone correspond only 
 to the flow between them, this backbone network can be viewed as an 
 isolated communication system where the routers create, delivery and
receive data packets. Therefore, the measures effectively correspond
to a small network handling a large amount of traffic and whith all
its nodes having a similar degree. For this reason, we are not able to
observe here the dependence with the node degree. However, at
variance with the analysis performed in \cite{flowalex}, we have
varied the length of the time windows used to extract the flux and its
deviation \cite{note2}. Once again, the results, depicted in Fig.\
\ref{fig3}, show that the exponent $\alpha$ is not universal and
radically depends on $M$. Note that, although the lower bound of
$\alpha=0.706>1/2$ is determined by the minimal resolution ($M=5$
minutes) of the raw data, further increasing $M$ will recover the
upper bound $\alpha=1$. 

In summary, in this paper we have derived a theoretical law 
for the dependence of fluctuations with the mean traffic
in a network. Such a dependence is governed by three factors: 
one related to the dynamics, one related to the topology,  
and one of statistical nature. More importantly, the theoretical law 
reveals that the previously claimed power-law scaling (with universal 
or non-universal exponents) has to be abandoned. Our numerical results and 
the analysis of real data confirm that, even in the  presence of correlations 
between packets, one cannot assume a single exponent to characterize 
the fluctuations of traffic for the whole network. Finally, we note that the 
scaling breakdown predicted here is amenable to experimental confirmation by measuring 
the traffic flow in large communication networks  so to capture the predicted 
(topological) effects of degree heterogeneity.

\begin{acknowledgments}
We thank A. Arenas, De Felice, G. Drovandi and J. Duch for helpful comments 
and discussions on this work. 
Y. M. is supported by MEC through the Ram\'{o}n y Cajal Program. 
This work has been partially supported by the Spanish DGICYT Projects 
FIS2006-12781-C02-01 and FIS2005-00337. 
\end{acknowledgments}


\begin{thebibliography}{99}
\bibitem{note1} We have checked that the behavior observed does not depend on the 
creation rate of packets, by considering uniform, exponential and power law distribution functions 
(in all cases, the distributions are characterized by the same average number 
of packets $p$).
\bibitem{note2} We are implicitly assuming that packets are uncorrelated. This approximation seems not to crude as the lifetime of packets in the Internet is at most of several seconds, while the minimal resolution of the raw data is 5 minutes. After all, the 5 minutes time window can also be considered as cumulative data. We have, however, redone the calculations taking nonconsecutive time windows to further avoid possible correlations with no qualitative changes in the results.
\end{thebibliography}
\end{document}